\documentclass[12pt]{elsart}  
\usepackage[dvips]{epsfig}  
  
\newenvironment{lyxlist}[1]  
{\begin{list}{}  
{\settowidth{\labelwidth}{#1}  
\setlength{\leftmargin}{\labelwidth}  
\addtolength{\leftmargin}{\labelsep}  
}}  
{\end{list}}  
  
\usepackage{amsmath,amssymb}  
\renewcommand{\d}{\,\mathrm{d}}  
\renewcommand{\vec}[1]{\boldsymbol{#1}}  
\newcommand{\displ}[1]{{\displaystyle #1}}  
  
\begin{document}  
  
\begin{frontmatter}  
\title{Renormalization group analysis of the quantum non-linear sigma model  
with a damping term}  
  
\author[torino]{Andrea Gamba\thanksref{mail}
} , 
\author[roma]{Marco Grilli\thanksref{mail}
}  
and \author[roma]{Claudio Castellani\thanksref{mail}
}

%\thanks[ccmail]{E-mail: castellani@roma1.infn.it}  
%\thanks[mgmail]{E-mail: grilli@roma1.infn.it}
%\thanks[agmail]{E-mail: gamba@polito.it}  

\thanks[mail]{E-mail: gamba@polito.it, grilli@roma1.infn.it, 
castellani@roma1.infn.it}

\address[torino]{Dipartimento di Matematica, Politecnico di Torino, 
10129 Torino, Italy%\\
%gamba@polito.it
%\\
%and INFN, Sezione di Milano, 20133 Milano, Italy 
}
  
\address[roma]{Istituto Nazionale di Fisica della Materia e Dipartimento
di Fisica, \\Universit\`a di Roma ``La Sapienza",  
Piazzale A.~Moro 2, 00185 Roma, Italy
%grilli@roma1.infn.it, castellani@roma1.infn.it
}

  
\begin{abstract}  
  
We investigate the behavior of the zero-temperature   
quantum non-linear sigma model in $d$ dimensions in the presence of
a damping term of the form \( f(\omega )\sim |\omega |^{\alpha } \),  
with \( 1\le \alpha <2 \). We find two fixed points:  
a spin-wave fixed point FP1 showing a dynamic scaling exponent  
\( z=1 \) and  
a dissipative fixed point FP2 with \( z>1. \) In the framework of the  
\( \epsilon  \)-expansion  
it is seen that there is a range of values 
\( \alpha_* (d) \le\alpha \le 2 \)  
where the point FP1 is  
stable with respect to 
%the dissipative fixed point 
FP2, so that the  
system realizes a \( z=1 \) quantum critical behavior even in the presence  
of a dissipative term. However, 
reasonable arguments suggest that in $d=2$ this range is very narrow.  
In the broken symmetry phase we discuss a phenomenological
scaling approach, treating 
damping as a perturbation of the ordered ground state.
The relation of these results with the pseudogap effect observed in   
underdoped layered cuprates is discussed.   
  
\end{abstract}  
\end{frontmatter}  
  
\section{Introduction}  
  
The presence of an antiferromagnetic (AF) phase rapidly replaced by an 
anomalous metallic phase upon doping is one of the prominent features of the
phase diagram of high temperature superconducting cuprates
\cite{johnston,rigamonti}.
Indeed it was early suggested \cite{anderson} that the proximity
to an insulating magnetic phase together with the nearly twodimensional
structure of these materials could be responsible for the
superconducting and anomalous normal-state  (i.e. non-Fermi liquid)
properties of these
systems via the creation of a quantum-disordered spin liquid
(resonating-valence-bond state). Since then a great deal of attention
has been devoted to the fascinating interplay between magnetism
and charge degrees of freedom, leading to various theoretical proposals
and to different scenarios for the metal-insulator transition 
at low doping.

One scenario is based on the tendency of a magnetically ordered
phase to segregate the additional holes due to doping
\cite{EKLmarder}. The tendency to phase separation is then frustrated by the
Coulombic repusion between the segregated carriers
\cite{emerykivelson}, thus leading to formation of hole-rich 
domain walls separating AF domains
in the form of stripe textures. 
Within this scenario, pseudogaps naturally arise in 
the underdoped phase of the materials
as a consequence of stripe fluctuations and local pair formation \cite{CDG}.
Obviously this will substantially affect the transition
to the AF insulating phase at low doping \cite{affleck,castroneto}. 

The transition between a magnetically ordered phase and a metallic
state can also be strongly affected by extrinsic ingredients
like the disorder induced by dopant ions (Sr or Zn in the case of
${\rm La_{2-x-y}Sr_xZn_yCuO_4}$). This may lead to the formation
of local random magnetic moments giving rise to spin-glass
ordering between the AF insulating and the paramagnetic metallic (PM)
phases \cite{cho92,chou95,gooding,aharony,johnston}. 

The stripe and the spin-glass scenarios clearly illustrate the
complicated and rich nature that the  AF-PM transition may take. 
However, other proposals are more directly related to 
magnetism and to the contiguity between the AF and the metallic phase.
These proposals start from the seminal work of Ref. \cite{chn89}
showing that the materials with half-filled CuO$_2$
planes are suitably described in terms of 
a twodimensional quantum Heisenberg AF model
with very small interplanar coupling. Following work 
\cite{sy92,cs93,csy94,sp93,sokol94} suggested that, although
the 2D model displays
long-range order at $T=0$, upon doping the added charges 
enhance the quantum spin fluctuations thus driving the
system into a disordered state even at zero temperature (quantum disordered
phase) \cite{sy92}. The occurrence of an AF
quantum critical point (QCP) provides a natural framework to intepret 
the scaling behavior at low temperatures and low frequencies
of the of the \( q \)-integrated magnetic susceptibility \( \chi (\omega )=  
\int \d ^{2} q \, \chi  
(\vec q ,\omega ) \).
The susceptibility \( \chi (\omega ) \) can be experimentally  
determined via magnetic  
resonance measurements of the nuclear copper spin-echo decay rate,  
\( T_{2G}^{-1} \) and spin-lattice relaxation rate, \( T_{1}^{-1} \)  
\cite{isyk93,rigamonti}. It is  
seen that in a range of temperatures  
\( T_{*}<T<T_{\mathrm{cr}} \), depending on doping,  
\( T_{1}T/T_{2G}=\mathrm{const} \) \cite{takigawa94}.  
Assuming the scaling relations \cite{sy92},   
 \( \chi (\vec q ,\omega )\sim \xi ^{2-\eta }\hat\chi  
 (|\vec q-\vec{Q}_\mathrm{AF}| \xi ,\omega /T) \) and  
 \( \xi \sim T^{-1/z} \), with \(z\) a dynamic critical exponent,  
 \( \eta  \) an anomalous dimension, \( \vec Q _{\mathrm{AF}} \)  
 the antiferromagnetic ordering  
wave-vector, \( \xi  \) the correlation length,  
\( \hat\chi  \) a scaling function, one gets  
\( T_{1}T/T_{2G}\sim T^{1-1/z} \) \cite{sp93,sokol94}. The constancy  
of this ratio is thus interpreted as the signature of a critical behavior  
corresponding to a \( z=1 \) value of the dynamic index  
\cite{sp93,sokol94,bpst94,bp95,cps96}.  
The strong critical spin fluctuations occurring at the
AF-QCP have also been claimed to provide a
possible source of pairing and normal-state anomalies.

Despite the apparent simplicity of a scenario involving
a direct transition between an insulating AF phase
and a PM phase, such a 
transition is far from being trivial and is still an open
problem. The main difficulty is in the lack of a microscopic
model being able to smoothly interpolate between 
an insulating AF ordered phase with gapped charge excitation
and a PM phase.
%, where low-energy particle-hole excitations
%open a channel for the damping of spin-waves.
On the one hand, the insulating AF phase is suitably represented 
by the Heisenberg model and, in particular, the long-wavelength and 
low frequency behavior of spin fluctuations in the CuO$_2$ planes   
can be well described by the   
quantum non-linear sigma model \cite{chn89}, which  
at zero temperature and in two spatial dimensions  
is characterized by the action  
\begin{equation}  
\label{pura}  
S=\frac{1}{2g}\int \d \omega \d ^{2}k  
\left( k^{2}+\frac{\omega ^{2}}{c^{2}}\right)  
\vec\phi _{\vec k ,\omega }\cdot \vec\phi _{-\vec k ,-\omega }  
\end{equation}  
Here \( \vec\phi  \) is a three-component  
vector field subject to the real-space   
constraint \( \phi _{\vec x ,\tau }^{2}=1 \) which describes  
the local staggered magnetization,  
and \(g\) is a coupling constant.   
  
It was first argued in \cite{sy92} that this description could be carried  
over to the case of the lightly doped compounds. In fact, one can  
think that in this case the \( \vec\phi  \) field continues  
giving an effective  
description of the electron spins localized on the Cu sites of the  
lattice, while the presence of itinerating holes   
provides a finite renormalization  
of the coupling constant.
The theory described by the action (\ref{pura}), besides having the correct  
symmetries, realizes a zero-temperature transition for a given critical  
value \( g_\mathrm{c} \) of the coupling (see \cite{chn89,haldane83} for the 
quantum version and  
\cite{polyakov75,bz76,zinn89} for the classical theory) thus providing  
a good model for the physical picture of the AF-QCP \cite{cs93}.  
 From this point of view the \( z=1 \) value of the dynamic scaling  
 exponent is a simple consequence of the Lorentz invariance of (\ref{pura}).   
  
However, it is clear that the non-linear $\sigma$ model 
does not properly describe the low-energy charge excitations,
which are characteristic of a metallic state possibly providing
a damping mechanism for spin-waves (see, e.g., case B of Ref. 
\cite{scs95}).
Therefore the model in Eq. \ref{pura} provides a good description
of a transition between an insulating (charge-gapped) AF (AFI) and an
insulating paramagnet (PI)., but it appears to be inappropriate
when spin-waves can decay in low-energy particle-hole excitations.

On the other hand, attempts have also been made to 
provide a microscopic basis for the transition to a PM phase. 
In this regard two distint cases A and B 
depending on the shape of the Fermi surface in the metal
were investigated in Ref. \cite{scs95} within a spin-density-wave
approach. In case A the AF wavevector
\({\vec Q}_\mathrm{AF}=(\pi/a,\pi/a)\) ($a$ is the lattice spacing)
was not able to connect points of the Fermi surface, so that 
momentum and energy conservation did not allow the decay of spin-waves
in particle-hole pairs. In this case the transition was naturally
characterized by a dynamic critical index $z=1$.
In case B, instead, there were (``hot'') points on the Fermi surface
connected by \({\vec Q}_\mathrm{AF}\) so that low-energy particle-hole pairs
could be created by the decay of low-energy spin-waves.
However, in both cases, A and B, the phase with AF long-range order 
is metallic and therefore the model does not suitably describe 
a transition between an AFI and a PM~\footnote{However, 
owing to the specific form of the spin-wave damping 
for a Fermi surface of type A, 
the non-linear $\sigma$ model would suitably describe the system also in
this case despite the metallic character of both the ordered
and disordered phases.}.
The case A above shares some common features with a mixed
spin-fermion model without a three-body (i.e. Kondo-like)
direct coupling between the local moments and the fermion spins
\cite{ss88}. This model has been shown \cite{sachdev94} to display 
a $T=0$ transition between a metallic (Fermi liquid) phase
with spin commensurate long-range order and a metallic (also 
Fermi liquid) quantum disordered phase. Owing to the absence
of the three body coupling responsible for the direct 
spin-wave decay into particle-hole pairs, also the QCP
of this model is characterized by $z=1$.

To the best of our knowledge no simple microscopic modelization
is presently available to describe a direct AFI-PM transition.
Of course the possibilities remain open that other extrinsic
mechanisms (stripes, disorder, ...) provide intermediate
steps between the AFI and the PM phases. Alternatively, one may also
envisage that an AFI to a PI phase transition occurs first at
lower doping, followed at slightly larger doping by a PI-PM
transition between two spin-disordered phases.

Together with the basic difficulty outlined above,
the additional question arises concerning the observed
$z=1$ critical behavior in the underdoped phase of the
cuprates and the way this can be reconciled with
the metallic character of this phase. In particular it
is believed that the Fermi surface of the cuprates in the metallic phase
contains ``hot'' points connected by the AF wavevector.
Therefore it should be possible for the spin-waves 
of arbitrarily low energy to decay into particle-hole pairs
and get overdamped. This would lead to a relaxational behavior
contrasting with the $z=1$ (i.e. propagating) critical
behavior of the spin-waves. Moreover, by accepting
a $z=2$ critical behavior in the metallic phase, it remains
to clarify how this may turn into the natural propagating
behavior, which characterizes the spin excitations in the
ordered side of the QCP.
This latter difficulty was also encountered and stressed
by the authors of Ref. \cite{scs95} for their case B: within their
mean-field description a $z=2$ behavior was obtained 
both in the disordered and the ordered AF (metallic) phases. This suggested
that the simple formulation B of the mean-field SDW model
was only applicable on the disordered side of (and not too close to)
the AF-QCP. 

In the present paper we will not attempt to find a microscopic 
foundation to a direct AFI-PM transition, but we will instead
start from a semiphenomenological point of view by introducing 
a field-theoretical generalization of the non-linear
$\sigma$ model action (\ref{pura})  to investigate
a) the condition to be fullfilled by a simple (\(\vec k\)-independent)
damping term in order to maintain a $z=1$ critical behavior;
b) how in the $z>1$ case an additional crossover energy must behave in the
ordered phase to provide a scale separating damped spin
excitations at high energy and propagating spin excitations 
at low energy.

The resulting theory will depend on the precise form of the  
effective damping term.  
 We first discuss the
simpler case of the metallic phase on the $g>g_c$ side of the QCP.
Within a microscopic model of spins coupled to free itinerant holes,  
integrating out the holes degrees of freedom in the  
random phase approximation gives a damping term of the form  
\( f(\omega )=\gamma |\omega | \) (see \em e.g.  
\em \cite{hertz76,scs95}). 
However,  
the choice of the adequate low frequency form of  
\( f(\omega ) \) should also  
take in consideration the fact that in underdoped cuprates a significant  
loss of low-frequency spectral weight is observed at low temperatures  
in the distribution of quasi-particles, a phenomenon 
often referred to as ``pseudogap''
(see \em e.g. \em \cite{rc97P}). The origin of this effect is a debated  
issue, and has been alternatively interpreted:
as the result of spin-fermion scattering  \cite{cms96}, 
as due to the intrinsecally
non-Fermi liquid nature of the system (Luttinger liquid) 
\cite{andersonlutt}, 
as a signal of the formation of preformed pairs for  
\( T>T_{\mathrm{c}} \) \cite{rc97P,ml96},
as due to stripe charge fluctuations \cite{CDG}.
In all cases, a precise computation of these effects implies solving  
a difficult self-consistency problem going well
beyond the RPA approach. We shall try here a rough,   
phenomenological approach, assuming that the damping coefficient  
\( \gamma  \) can acquire at low frequencies an \( \omega  \)-dependence  
\( \sim |\omega |^{x} \), thus giving rise to an effective damping  
term of the form \( f(\omega )\sim |\omega |^{\alpha } \), with  
\( \alpha =1+x>1 \)~\footnote{The $\alpha=1$ case 
corresponds to the case considered in 
Ref. \cite{scs95} within a large-N framework.}.
  
Once chosen a damping term \(f(\omega)\)   
we perform a renormalization group (RG) analysis of the resulting  
theory in the proximity of the QCP using the 
mo\-men\-tum-shell method and \( \epsilon  \)-expansion.  
We observe the existence of two fixed points,  
which we shall denote FP1 and FP2, showing respectively a  
\( z=1 \), undamped spin-wave  
behavior and a \( z>1 \), dissipative behavior. The point FP1 corresponds  
to the zero-temperature critical point of a pure quantum non-linear  
sigma model without any damping term, but it becomes unstable in a  
wide region of the parameters with respect to the dissipative fixed  
point FP2.  
  
There exists a range of values \( \alpha_*\le\alpha \le 2 \)  
for which one obtains a \( z\equiv 1 \) dynamic  
scaling exponent in the proximity of the stable quantum critical point,  
even in the presence of a damping term. In other words, there are  
``soft'' damping terms which do not destroy the \( z=1 \),  
spin-wave behavior. Here \( \alpha_* \) is the 
dimension dependent lower bound for  
the exponents in the soft damping term, which we determine to be  
\( \alpha_* =2-\eta(\alpha_*)=2-\eta_1 \), 
where \( \eta(\alpha) \) is the anomalous field 
dimension and $\eta_1$ is its value at the FP1.  
It is worth noting that this result bears
a resemblance to what is found in classical models with long-range 
interactions~\cite{fmn72,sak73}. 
In particular the scaling exponent 
of the damping stems from the non-analytic
form of $f(\omega)$, which is not reconstructed by the
RG corrections (for a discussion of this point in classical
models see Ref.~\cite{aharony76}). 
Therefore $\gamma$ does not acquire singular 
corrections but for the anomalous field dimensions.
 
Within the framework of an expansion in  
\( \epsilon =d+z^{(0)}-2 \), where \( d\) is the spatial  
dimension of the system and \( z^{(0)} \) is the bare  
value of the dynamic scaling exponent in the vicinity of the stable  
fixed point, one would get 
\( \alpha_*=2-(\alpha\epsilon_2)/(1+\alpha)=2-\epsilon_1 \),
where the 1 or 2 subscripts in $\epsilon$ refer to FP1 or FP2
respectively.
However,  the extrapolation of results valid for  
\( \epsilon \simeq 0 \) to the  
physical region $d=2$ should be done with great care. In particular, using  
the numerical results for the critical exponents of the  
\( \mathrm{O}(3) \) model in 2+1 dimensions, it appears that  
\( \alpha_* \) is very near to two and   
the range of ``soft'' \( \alpha  \) values should  
be a very narrow one. This means that within the here-considered model
the appearance of a \( z=1 \) critical  
behavior in the cuprates
cannot be accounted for in the effective framework without  
assuming a value \( \alpha \approx 2 \), \em i.e.\em ,  
an almost linear decay of the quasi-particle  
spectral weight for low frequencies. This spectral weight
suppression is a rather
severe condition to be obtained from the magnetic scattering itself,
as also suggested by a direct perturbative
calculation within a spin-fermion model \cite{chubukov}, where
density of states and vertex corrections to the fermion bubble
produced only a minor change of the power-law dependence of the
damping term.
It is worth noting that a substantial
spectral weight reduction is however consistent with Anderson's
idea that a smooth connection between a charge-gapped insulator
and a Fermi-liquid metal is hardly conceivable.
  
As far as the ordered phase is concerned, (point (b) above) 
it is necessary to
take into account the presence of a  non-zero staggered magnetization  
\( N_{0} \), which is accompanied by the 
presence of two quasi-particle bands
separated by an energy \( \Delta \).  
This suggests \cite{scs95} to introduce
a damping term having two distinct asymptotics:
\(f(\omega)\sim|\omega|^\alpha \) for \(|\omega|\gg\Delta\) and 
\(f(\omega)\sim\omega^b\) for \(|\omega|\ll\Delta\), with
\(1\le\alpha<2\le b<3\)~\footnote{In principle the damping  
could also depend on \( \vec q =\vec Q _{\mathrm{AF}}+\vec k  \)  
in the ordered (as well as in the disordered) phase,
but in a first approximation we ignore this dependence 
mainly for simplicity reasons, but also because in the insulating
phase, where a gap for the charge excitations providing
spin-wave decay is present, 
it would be unnatural to expect a relevant q-dependence.}
The difficult point here is that the scale \( \Delta  \) separating 
damped and undamped spin excitations vanishes at the  
transition.
This prevents using a conventional RG approach, and we will
resort to simple scaling arguments and phenomenological assumptions.
Our analysis does not attain the same degree of 
reliability of a traditional Landau-Ginzburg-Wilson theory
and has a heuristic character.
We shall assume that \(\Delta\) has a power-law behavior, to be
determined by matching the hydrodynamic behaviors on the
two sides of the quantum transition.
For \(\alpha<\alpha_*\), \em i.e.\em for $z>1$, 
we find that (i) \(\Delta\) should close at the ``microscopic"
scale as \(\xi^{-z}\) for \(\xi\to\infty\),
where \(\xi\) is a correlation length,
and that (ii) 
the spin-wave velocity \(c_\mathrm{sw}\)
vanishes
as \(\xi^{1-z}\), to be consistent with the presence of damping.

The paper is organized as follows: in Sec.  
\ref{rengreq} we derive the RG equations  
for the zero-temperature quantum non-linear sigma model with a damping  
term of generic form, following the scheme exposed in \cite{chn89,np77}.  
This part is rather technical and it can be skipped in a first reading  
without loosing the general meaning of the paper.  
  
In Sec. \ref{fixpanrgeq} we perform  
the fixed point analysis of the RG equations for  
a damping term \( f(\omega )=\gamma |\omega |^{\alpha } \).  
Here we show the existence of the two fixed points mentioned above,  
study their respective regions of stability and compute the relevant  
critical exponents, with particular regard to the value of the dynamic  
scaling exponent \( z \).  
  
In Sec. \ref{ordphase} we discuss damping in the ordered phase,
introducing a phenomenological 
scale \(\Delta\) vanishing at the quantum critical
point with a critical exponent that we determine by matching the
hydrodynamic behaviors on the two sides of the transition.
We also compute the critical behavior of the spin-wave velocity
\(c_\mathrm{sw}\) in the presence of damping. 
  
In 
Sec. \ref{conclusions} we present some conclusive remarks 
and discuss the connection between our model and other physically
interesting systems, like macroscopic quantum tunnelling and 
spin chains with long range interaction.

\section{Renormalization group equations}  
  
\label{rengreq}Let us consider the Euclidean action of the  
quantum non-linear sigma  
model \cite{chn89,haldane83,polyakov75,bz76} at zero temperature  
in \( d \) spatial dimensions, with a damping  
term \( f(\omega ) \) of generic form and  
in the presence of a constant magnetic field \( h \):   
\begin{equation}  
\label{azione1}  
{S}=\frac{1}{2g}\int \d \kappa \left( k^{2}+  
\frac{\omega ^{2}}{c^{2}}+f(\omega )\right) \,  
\vec\phi _{\vec\kappa }\cdot \vec\phi _{-\vec\kappa }-h\phi _{\vec 0}^{N}  
\end{equation}  
Here \( \vec\phi  \) is an \( N \)-component  
field satisfying the constraint \(\vec\phi _{x,\tau }^{2}=1 \)  
in real space, \( \vec\kappa \equiv (\vec k ,\omega ) \),  
\( \d \kappa \equiv (2\pi )^{-(d+1)}\d ^{d}k\d \omega  \).  
The magnetic field is oriented along  
the \( N \)-th direction.
The momentum and frequency integrations have been rescaled  
so that the adimensional spin velocity  \(c \)  
has bare value equal to one and 
the momentum integration is cut off at  
\( \Lambda \equiv 1 \).
  
With our rescaling, in (\ref{azione1}) the dimensionless  
coupling constant \( g \) is related to the bare  
(dimensional) spin-wave velocity \( c_{0} \), the  
physical cut-off \( \Lambda  \) and the bare spin-stiffness constant  
\( \rho _{0} \) by the relation  
\( g=\hbar c_{0}\Lambda ^{d-1}/\rho _{0} \) \cite{chn89}.  
In the large \( S \) (here $S$ is the spin) limit  
 \( \rho _{0} \) and \( c_{0} \) are related to the  
 coupling constant of the Heisenberg model  
\( J \) and the lattice spacing \( a \) by \( \rho _{0}=JS^{2}a^{2-d} \)  
and \( c_{0}=2\sqrt{d}JSa/\hbar  \) \cite{haldane83}  
with \( \Lambda \sim a^{-1} \). \(h\) is also adimensional.  
  
Action (\ref{azione1}) is formally equivalent to the  
Hamiltonian of an anisotropic corresponding  
statistical mechanical model in \( d+1 \) dimensions.  
The critical point \(g=g_*\) of the classical model is here reinterpreted  
as a zero-temperature quantum critical point, separating the ordered  
and the disordered ground state of the system, which are realized,  
respectively, for \( g<g_{*} \) and \( g>g_{*} \).  
In order to study the long-distance, low-frequency behavior of the  
\( \vec\phi  \) field in the proximity of the  
quantum critical point we shall use  
the renormalization group and the \( \epsilon  \)-expansion,  
following Refs. \cite{chn89,np77}.  
  
In the broken symmetry phase we let  
\( \vec\pi =(\phi ^{1},\ldots ,\phi ^{N-1}) \), \( \sigma =\phi ^{N} \).  
Keeping only terms up to  
\( O(g^{2}) \), action (\ref{azione1}) takes the form   
\begin{eqnarray}  
S & = & \phantom{+}\frac{1}{2g}\int \d \kappa \, \Delta _{(f)}(k,\omega )\,  
\vec\pi _{\vec\kappa }\cdot \vec\pi _{-\vec\kappa }\nonumber \\  
 &  & +\frac{1}{2g}\int \d ^{(4)}\kappa  
 \left( -\vec k _{1}\cdot \vec k _{3}-\frac{\omega _{1}\omega _{3}}{c^{2}}+  
 \frac{1}{4}f(\omega _{1}+\omega _{2})\right) \vec\pi _{\vec\kappa _{1}}  
 \cdot \vec\pi _{\vec\kappa _{2}}\, \vec\pi _{\vec\kappa _{3}}  
 \cdot \vec\pi _{\vec\kappa _{4}}\nonumber \\  
\label{azione2}  &  & +\frac{h}{8}\int \d ^{(4)}\kappa  
\, \, \vec\pi _{\vec\kappa _{1}}\cdot \vec\pi _{\vec\kappa _{2}}  
\, \vec\pi _{\vec\kappa _{3}}\cdot \vec\pi _{\vec\kappa _{4}}\\  
 &  & -\frac{{V} }{2}\int \d \kappa \, \vec\pi _{\vec\kappa }  
 \cdot \vec\pi _{-\vec\kappa }\, +\, O(g^{3})\nonumber  
\end{eqnarray}  
where the inverse propagator, given by  
\[  
\Delta _{(f)}(k,\omega )=k^{2}+\frac{\omega ^{2}}{c^{2}}+f(\omega )+gh\]  
depends on the damping term \(f(\omega)\), and  
the measure term \( \d ^{(2n)}\kappa  \) is  
\[  
\d ^{(2n)}\kappa =(2\pi )^{d+1}\delta ^{(d+1)}(\vec\kappa _{1}+  
\cdots +\vec\kappa _{2n})\d \kappa _{1}\cdots \d \kappa _{2n}\]  
In (\ref{azione2}) the term proportional to the extended phase-space volume  
\begin{equation}  
\label{rsd}  
{V}  =\frac{S_{d}}{(2\pi )^{d+1}}\frac{2\Omega }{d}=  
\int _{-\Omega }^{\Omega }\frac{\d \omega }{2\pi }  
\int _{k<1}\frac{\d ^{d}k}{(2\pi )^{d}}=\int \d \kappa  
\end{equation}  
comes  
from the perturbative evaluation of the Jacobian   
\begin{equation}  
\label{pxt}  
\prod _{x,\tau }\frac{1}{\sqrt{1-\pi ^{2}_{x,\tau }}}  
\end{equation}  
In (\ref{rsd}) \( \Omega  \) is an arbitrary  
frequency cut-off which does not enter the final result,  
and \( S_{d}=2\pi ^{d/2}/\Gamma (d/2) \).  
  
Let us denote the vertices appearing (in the same order) in  
(\ref{azione2}) as  
  
\vspace{0.30cm}  
{\centering \epsfig{file=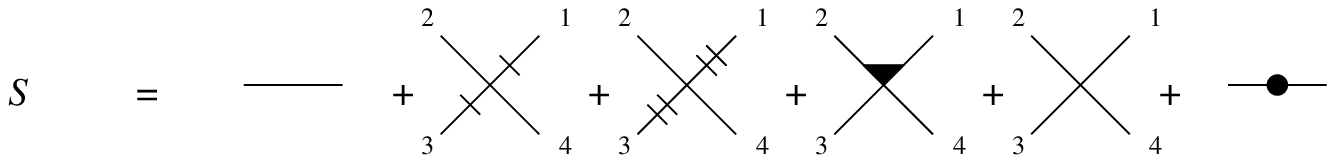} \par}  
\vspace{0.30cm}  
  
\noindent where the cut, double cut and triangle indicate respectively  
a \( \vec k  \), \( \omega  \) and \( f(\omega ) \) insertion,  
and the dotted line stays for the first perturbative  
term of the Jacobian. Let then \( S=S_{0}+S_{1} \), with   
\[  
S_{0}=\frac{1}{2g}\int \d \kappa \, \Delta _{(f)}(k,\omega )\,  
\vec\pi _{\vec\kappa }\cdot \vec\pi _{-\vec\kappa }\]  
We shall use the momentum-shell  
method to derive recursion relations \em \`a la \em Wilson \cite{chn89,np77},  
integrating out the \( \vec\pi _{\vec\kappa } \) with  
\( e^{-l}<k<1 \), \( l\simeq 0 \) and all \( \omega  \), and rescaling  
\[  
\vec k \rightarrow e^{-l}\vec k ,\, \, \, \omega  
\rightarrow e^{-zl}\omega ,\, \, \, \vec\pi \rightarrow \zeta \vec\pi \]  
hereby introducing  
a dynamic scaling exponent \( z \) and a wave  
function renormalization \( \zeta  \).  
We average with respect to the following propagator:  
\[  
\langle \pi ^{i}_{\vec\kappa }\pi _{\vec\kappa '}^{j}  
\rangle _{l}=\left\{ \begin{array}{ll}  
\displ{ (2\pi)^{d+1}\frac{\delta ^{(d+1)}(\vec\kappa +  
\vec\kappa ')\delta ^{ij} g }{\Delta _{(f)}(k,\omega )}} & ,  
\, \, \text {for}\, \, \, e^{-l}<k<1,\, \, e^{-l}<k'<1\\  
\pi ^{i}_{\vec\kappa }\pi ^{j}_{\vec\kappa '}  
\phantom {\Big (} & ,\, \, \text {for}\, \, \, k<e^{-l},\, \, k'<e^{-l}\\  
0 & ,\, \, \text {otherwise}  
\end{array}  
\right. \]  
The first perturbative term of the Jacobian can be written as  
\begin{equation}  
\label{jacobian}  
-\frac{{V}  }{2g}\int _{k<1}\d \kappa \,  
\vec\pi _{\vec\kappa }\cdot \vec\pi _{-\vec\kappa }=  
-\frac{1}{2g}\left[ \int _{k<e^{-l}}\d \kappa +\int _{e^{-l}<k<1}\d  
\kappa \right] \int _{k<1}\d \kappa \, \vec\pi _{\vec\kappa }  
\cdot \vec\pi _{-\vec\kappa }  
\end{equation}  
and we will indicate it pictorially as   
  
\vspace{0.30cm}  
{\centering \epsfig{file=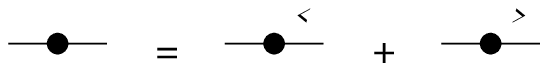} \par}  
\vspace{0.30cm}  
  
\noindent After integration on the momentum shell the first term represents  
the modified first-loop Jacobian contribution, while the second has  
to be absorbed in the mass renormalization and cancels the  
contribution from the second vertex in Eq. (\ref{azione2})  
(see below),  
thus maintaining the theory massless at \(h=0\).  
  
The usual diagrams renormalizing the \( k^{2} \) and \( \omega ^{2} \)  
terms both give a \( \frac{1}{2}glI_{(f)} \) contribution,  
with  
\begin{equation}  
\label{loop0}  
I_{(f)}=\frac{S_{d}}{(2\pi )^{d+1}}\int _{-\infty }^{+\infty }  
\frac{\d \omega }{\Delta _{(f)}(1,\omega )}  
\end{equation}  
In order to renormalize \( f \) and \( h \) we have to consider the  
following contributions:  
  
\vspace{0.30cm}  
{\centering \epsfig{file=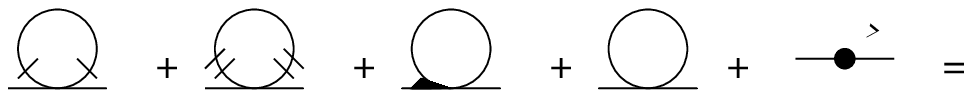} \vskip -1.8em\par}  
\vspace{0.30cm}  
  
\begin{eqnarray}  
 \label{diagrammi}  & = & \frac{1}{2}  
 \int _{k_{1}<e^{-l}}\d \kappa _{1}\int _{e^{-l}<k_{2}<1}\d \kappa _{2}  
 \cdot \\  
\nonumber &  & \cdot \left[  
\frac{(k_{2})^{2}}{\Delta _{(f)}(k_{2},\omega _{2})}+  
\frac{(\omega _{2}/c)^{2}}{\Delta _{(f)}(k_{2},\omega _{2})}+  
\frac{f(\omega _{1}+\omega _{2})}{\Delta _{(f)}(k_{2},\omega _{2})}+
\frac{1}{2}
\frac{(N+1)gh}{\Delta _{(f)}(k_{2},\omega _{2})} -1\right]  
\vec\pi _{\vec\kappa _{1}}\cdot \vec\pi _{-\vec\kappa _{1}}\\  
\nonumber & \simeq  & \frac{1}{2}  
\int _{k_{1}<e^{-l}}\d \kappa _{1}\left[ 
\int _{e^{-l}<k_{2}<1}\!\!\frac{f(\omega _{1}+\omega _{2})  
-f(\omega _{2})}{\Delta _{(f)}(k_{2},\omega _{2})}\d \kappa _{2}+
\frac{N-1}{2}ghlI_{(f)}   \right]    
\vec\pi _{\vec\kappa _{1}}\cdot \vec\pi _{-\vec\kappa _{1}}  
\end{eqnarray}  
The symmetry of (\ref{azione1}) requires that \( h \) scales as  
\( h\rightarrow \zeta h \). Since \( h \) plays  
in (\ref{azione2}) the role of a mass, from one-loop mass  
renormalization we get  
that the field rescaling factor is  
\[  
\zeta \simeq \exp \left[ l\left( d+z-\frac{N-1}{2}gI_{(f)}\right)  
\right] \]  
Defining \( \hat f (\omega )=\omega ^{^{2}}/c^{2}+f(\omega ) \)  
we finally get the RG equations:  
\begin{eqnarray}  
\label{rtildeg} \frac{\d g}{\d l} & = & \left[ 2-d-z+(N-2)gI_{(f)}\right] g\\  
\label{rhatf} \frac{\d\hat f }{\d l}(\omega ) & = &  
\left( 2-gI_{(f)}-z\omega \frac{\d}{\d\omega }\right)  
\hat f (\omega )+gC_{(f)}(\omega )  
\end{eqnarray}  
with  
\begin{equation}  
\label{conv0}  
C_{(f)}(\omega )=\frac{S_{d}}{(2\pi )^{d+1}}  
\int _{-\infty }^{+\infty }\frac{\hat f (\omega '+\omega )-  
\hat f (\omega ')}{\Delta _{(f)}(1,\omega ')}\d \omega '  
\end{equation}  
Equation (\ref{rtildeg}) results as usual from field rescaling  
and renormalization  
of the \( k^{2} \) term. The term \( z\omega \frac{\d}{\d\omega } \)  
in Eq. (\ref{rhatf}) describes the frequency rescaling  
and \( -gI_{(f)} + g C_{(f)} \)  
is related to the contributions in Eq. (\ref{diagrammi}).  
  
Let us now consider a damping term of the form \( f(\omega )=  
f_{1}(|\omega |) \), with \( f_{1} \) a generic differentiable  
function behaving as \( f_{1}(\omega )\sim \omega ^{a} \) for  
\( \omega \rightarrow +\infty  \). One observes that for \( a<3 \)  
and \( f_{1}'' \) integrable  
in a neighborhood of \( \omega =0 \) the term  
\( C_{(f)}(\omega ) \) is twice continuously differentiable,  
so that for \( \omega \simeq 0 \) it is possible to expand  
\begin{equation}  
\label{sviluppo}  
C_{(f)}(\omega )=\left( \frac{1}{c^{2}}I_{(f)}+2K_{(f)}\right)  
\omega ^{2}+\text {higher\, order\, terms}  
\end{equation}  
with   
\begin{equation}  
\label{kappaf}  
K_{(f)}=\frac{1}{2}\frac{S_{d}}{(2\pi )^{d+1}}\left[ f_{1}'(0)+  
\int _{0}^{+\infty }\frac{f_{1}''(\omega )}{1+\omega ^{2}/c^{2}+  
f_{1}(\omega )}\d \omega \right]  
\end{equation}  
In order to prove (\ref{sviluppo},\ref{kappaf})  
it is enough to rewrite (\ref{conv0}) as  
\begin{eqnarray*}  
\nonumber
C_{(f)}(\omega)&=&  
\frac{S_d}{(2\pi)^{d+1}}  
\int _{0}^\omega 
\frac{f_{1}(\omega '+\omega )+f_{1}(\omega -\omega' )-  
2f_{1}(\omega ')}{1+{\omega '}^{2}/c^{2}+f_{1}(\omega ')}\d \omega ' \\ 
\nonumber
&+&\frac{S_d}{(2\pi)^{d+1}}  
\int _{\omega }^{\infty }
\frac{f_{1}(\omega '+\omega )+f_{1}(\omega' -\omega )-  
2f_{1}(\omega ')}{1+{\omega '}^{2}/c^{2}+f_{1}(\omega ')}\d \omega '  
\end{eqnarray*}  
and to differentiate the expression twice with respect to \(\omega\).  
\( C_{(f)}''(\omega ) \) is continuous but not differentiable in  
\( \omega =0 \).  
  
The content of Eq. (\ref{sviluppo}) is that  
\em the non analyticity of the damping term  
in \( \omega =0 \)
is not reproduced under  
renormalization\em. This implies that  the \( f(\omega ) \)  
term renormalizes only  
according to its bare dimensions and to the contribution of the   
field rescaling factor. On  
the other hand,   
\(f(\omega)\)  
contributes to the renormalization of the  
\( \omega ^{2} \) term.  
  
According to Eq. (\ref{kappaf}),  
\( K_{(f)} \) is essentially a measure of the non linearity  
of \( f \) and is not strongly  
dependent on its explicit form.   
  
For \( f(\omega )=\gamma |\omega |^{\alpha } \), with  
\( 1\le \alpha <2 \) and \( \gamma  \) a damping coefficient, equation  
(\ref{rhatf}), together with (\ref{sviluppo},\ref{kappaf}), gives   
\begin{eqnarray}  
\label{grt2} \frac{\d\gamma }{\d l} & = & [2-  
\alpha z-gI_{\alpha }(c,\gamma )]\gamma \\  
\label{grt3} \frac{\d}{\d l}\frac{1}{c^{2}} & = & 2(1-z)\frac{1}{c^{2}}+  
2gK_{\alpha }(c,\gamma )  
\end{eqnarray}  
with \em   
\begin{eqnarray}  
\label{loop} I_{\alpha }(c,\gamma ) & = & \frac{S_{d}}{(2\pi )^{d+1}}  
\int _{-\infty }^{+\infty }\frac{\d \omega }{1+\omega ^{2}/c^{2}+  
\gamma |\omega |^{\alpha }}\\  
\label{conv} K_{\alpha }(c,\gamma ) & = & \frac{S_{d}}{(2\pi )^{d+1}}  
\frac{\gamma }{2}\alpha (\alpha -1)\int _{0}^{+\infty }  
\frac{\omega ^{\alpha -2}\d \omega }{1+\omega ^{2}/c^{2}+  
\gamma \omega ^{\alpha }}  
\end{eqnarray}  
\em In Appendix \ref{compcritind} the limit for \( \alpha \rightarrow 1 \)  
of these expression  
is discussed (see (\ref{loop2},\ref{conv2})).

\section{Fixed point analysis of the RG equations}  
\label{fixpanrgeq}
\subsection{One-loop analysis}  
  
We shall now study the RG equations obtained  
in the previous section  
for the zero-temperature critical point of the quantum
non-linear sigma model with a damping term \( f(\omega )=  
\gamma |\omega |^{\alpha } \), described by the action  
(\ref{azione1}). 
For the sake of clarity we rewrite here the  
RG equations (\ref{rtildeg},\ref{grt2},\ref{grt3}):   
\begin{eqnarray}  
\label{grt1_} \frac{\d g}{\d l} 
& = & [2-d-z+g(N-2)I_{\alpha }(c,\gamma )]g\\  
\label{grt2_} \frac{\d\gamma }{\d l} & = & [2-\alpha z-  
gI_{\alpha }(c,\gamma )]\gamma \\  
\label{grt3_} \frac{\d}{\d l}\frac{1}{c^{2}} & = & 2(1-z)  
\frac{1}{c^{2}}+2gK_{\alpha }(c,\gamma )  
\end{eqnarray}  
where the integrals \( I_{\alpha } \) and \( K_{\alpha } \) were  
defined in (\ref{loop},\ref{conv}). The dynamic scaling  
exponent \( z \) was introduced when choosing to rescale  
frequencies as  
\( \omega \rightarrow e^{-zl}\omega  \), \( l \) being the  
parameter of the renormalization group.  
  
We look for fixed points of the RG transformation  
(\ref{grt1_}-\ref{grt3_}). One immediately  
finds two fixed points, characterized by  
\begin{lyxlist}{00.00}  
  
\item [FP1:]\( g_{*,1}=\displ {\frac{d-1}{(N-2)\pi c_{*}}},\, \, \,  
\gamma _{*}=0,\, \, \, c_{*}=\mathrm{const},\, \, \, z_{1}=1 \),~~~  
for \( d>1 \);  
  
\item [FP2:]\( g_{*,2}=\displ {  
\frac{\alpha d-2\alpha +2}{[(N-2)\alpha +1]I_{\alpha }(c_{*},  
\gamma _{*})}},\, \, \, c^{\alpha }_{*}\gamma _{*}=u_{*},\, \, \,  
z_{2}=\displ {\frac{2N-2-d}{(N-2)\alpha +1}} \),\\  
for \(  d >2- \displ{\frac{2}{\alpha}} \); \noindent  
  
\end{lyxlist}  
where \( u_{*} \) is the solution of  
\begin{equation}  
\label{vincolo}  
\tilde I _{\alpha }(u)=
\frac{\alpha d-2\alpha +2}{2N-3-d-(N-2)\alpha }  
\tilde K _{\alpha }(u)  
\end{equation}  
with \( \tilde I _{\alpha }(u)\equiv c^{-1}I_{\alpha }(c,uc^{-\alpha })=  
I_{\alpha }(1,u) \), \( \tilde K _{\alpha }(u)  
\equiv cK_{\alpha }(c,uc^{-\alpha })=K_{\alpha }(1,u) \).  
  
Strictly speaking, FP1 and FP2 are two \em lines \em of  
fixed points, since in both cases  
\( c_{*} \) is arbitrary. In the case of FP2 either  
\( c_{*} \) or \( \gamma _{*} \) can  
be arbitrarily chosen, and the remaining one is then determined through  
the constraint \( c^\alpha_{*}\gamma _{*}=u_{*} \).  
  
Concerning Eq. (\ref{vincolo}), note that, letting
\(\mu=\frac{\alpha d-2\alpha+2}{2N-3-d-(N-2)\alpha}\):   
\begin{enumerate}  
  
\item a unique solution \( u_{*} \) exists for  
\( \mu>0 \);  
  
\item \( u_{*}\rightarrow +\infty  \) for  
\( \mu\rightarrow 0 \);  
  
\item no solution exists for \( \mu<0 \).  
  
\end{enumerate}  
Together with \( d>2-2/\alpha \), this implies 
\( d< 2N-3-(N-2)\alpha \).  
These are all consequences of the following properties:  
\( \tilde I _{\alpha }(u) \) is a monotonically decreasing  
function of \(u\), tending to a finite value for  
\( u\rightarrow 0 \) and vanishing for  
\( u\rightarrow +\infty  \); \( \tilde K _{\alpha }(u) \)  
is an increasing function, vanishing for \( u\rightarrow 0 \)  
and unbounded for  
\( u\rightarrow +\infty  \).  
%These properties are easily  
%verified (analytically for \(\tilde I_\alpha\) and numerically  
%for \(\tilde K_\alpha\) with \(\alpha>1\)).  
  
The analogy with the statistical mechanical model shows that the  
perturbative  
expansion is in the parameter \( \epsilon =d+z^{(0)}-2 \), where  
\( z^{(0)} \) is the value of the dynamic  
scaling exponent obtained from purely dimensional considerations.  
For the \( \gamma =0 \) case Lorentz invariance imposes \( z^{(0)}=1 \),  
 while in the presence  
of a non-zero damping term \( \sim |\omega |^{\alpha } \) with  
\( \alpha  \) not too close to \( 2 \) we expect the  
latter to be more relevant than \( \omega ^{2} \), and  
\( z^{(0)}=2/\alpha  \). Letting then \( \delta z\equiv z-z^{(0)} \)  
be the one-loop  
perturbative correction to the bare dynamic exponent \( z^{(0)} \)  
and \( \eta  \) the field  
anomalous dimension, one finds in the two cases :  
\begin{lyxlist}{00.00}  
  
\item [FP1:]\( z_{1}^{(0)}=1 \),~~ \( \delta z_{1}=0 \),~~ \(  
\epsilon _{1}=d-1 \),~~ \( g_{*,1}=\displ  
{\frac{\epsilon _{1}}{(N-2)\pi c_{*}}} \),~~ 
\( \eta _{1}=\displ{\frac{\epsilon _{1}}{N-2}} \)  
  
\item [FP2:]\( z_{2}^{(0)}=\displ  
{\frac{2}{\alpha }} \),~~ \( \delta z_{2}=\displ  
{-\frac{\epsilon _{2}}{(N-2)\alpha +1}} \),~~ 
\( \epsilon _{2}=d+  
\displ {\frac{2}{\alpha }}-2 \), \\  
\( g_{*,2}=\displ {\frac{\alpha }{(N-2)\alpha +1}  
\frac{\epsilon _{2}}{I_{\alpha }(c_{*},\gamma _{*})}} \),  
~~ \( \eta _{2}=\displ {\frac{\alpha \epsilon _{2}}{(N-2)\alpha +1}} \), \\  
with \( c^{\alpha }_{*}\gamma _{*}=u_{*} \) the solution of  
\( \tilde I _{\alpha }(u)=\displ {\frac{\alpha }{(N-2)\alpha +1}  
\frac{\epsilon _{2}}{z_{2}-1}}\tilde K _{\alpha }(u) \).  
  
\end{lyxlist}  
The point FP2 exists for \( \epsilon _{2}>0 \).  
For \( \epsilon _{2}<0 \) one gets the unphysical situation  
\( g_{*,2}<0 \). The additional condition \(d< 2N-3-(N-2)\alpha \)  
amounts to \( z_2 = \frac{2}{\alpha}- 
\frac{\epsilon_2}{(N-2)\alpha+1}>1\),  
which is the condition of stability to order \(\epsilon_2 \)  
of FP2 with respect to FP1. More precisely, at \(z_2=1 \)  
FP2 merges into FP1 (see below).  
For \( \epsilon _{2}\rightarrow 0 \)  
one has \( c^{\alpha }_{*}\gamma _{*}\rightarrow +\infty  \);  
keeping \( \gamma _{*} \) fixed this  
gives \( c_{*}\rightarrow +\infty  \), as expected,  
since for FP2 one has \( 1/c_{*}^{2}=0 \) at 0-th order  
in \( \epsilon _{2} \). For \( \alpha >1 \), (\ref{loop})  
shows that \( I_\alpha (+\infty ,\gamma ) \) is  
finite, so that \( g_{*,2}=O(\epsilon _{2}) \). The same  
is true in the \( \alpha =1 \) case, but some care  
must be exerted. A direct computation using (\ref{loop2},\ref{conv2})  
leads to ambiguous  
results, since in this case the frequency--wave-vector integration  
shell becomes singular: this is discussed more at length  
in Appendix \ref{compcritind}. We just observe here that for  
\( \alpha \rightarrow 1 \), \( \epsilon _{2}\rightarrow 0 \)  
implies \( d\rightarrow 0 \) and  
one has \( I_\alpha(+\infty ,\gamma )\sim \frac{d}{\alpha -1} \).  
One can send \( \epsilon _{2}\rightarrow 0 \) either with  
fixed \( \alpha \ne 1 \), or with fixed \( d\ne 0 \), and  
in both cases the loop integral \( I_{\alpha }(c_{*},\gamma _{*}) \)  
is finite at \( O(\epsilon _{2}) \).  
  
In order to compute the critical exponent \( \nu  \) we linearize the RG  
equations (\ref{grt1_}-\ref{grt3_}) and find the eigenvalues  
of the linearized transformation (see Appendix \ref{compcritind} 
for more detail):   
\begin{lyxlist}{00.00}  
  
\item [FP1:]\( \displ {\frac{1}{\nu _{1}}}  
\equiv \omega _{1}^{\mathrm{R}}=\epsilon _{1} \),~~~ \( 0 \),~~~\(  
\omega _{1}^{\mathrm{A}}=2-\alpha -
\displ{\frac{\epsilon _{1}}{N-2}} \) ~~~ for \( d>1 \);  
  
\item [FP2:]\( \displ {\frac{1}{\nu _{2}}}\equiv  
\omega _{2}^{\mathrm{R}}=\epsilon _{2} \),~~~ \( 0 \),~~~  
\( \omega _{2}^{\mathrm{A}}=\displ {2-\frac{4}{\alpha }+  
\frac{2}{(N-2)\alpha+1 }\epsilon _{2}} \)~~~\\ for  
\(d>2 -2/\alpha ,\, \,\,
2N-3-d-(N-2)\alpha>0 \).  
  
\end{lyxlist}  
This shows that the point FP2 always has one attractive direction  
in its region of existence, while \( \omega_1^\mathrm{A} >0 ~(<0) \)  
depends on \(2N-3-d-(N-2)\alpha<0~(>0)\).  
\begin{figure}  
{\setlength{\unitlength}{0.009\textwidth}\centering  
\bigskip
\epsfig{file=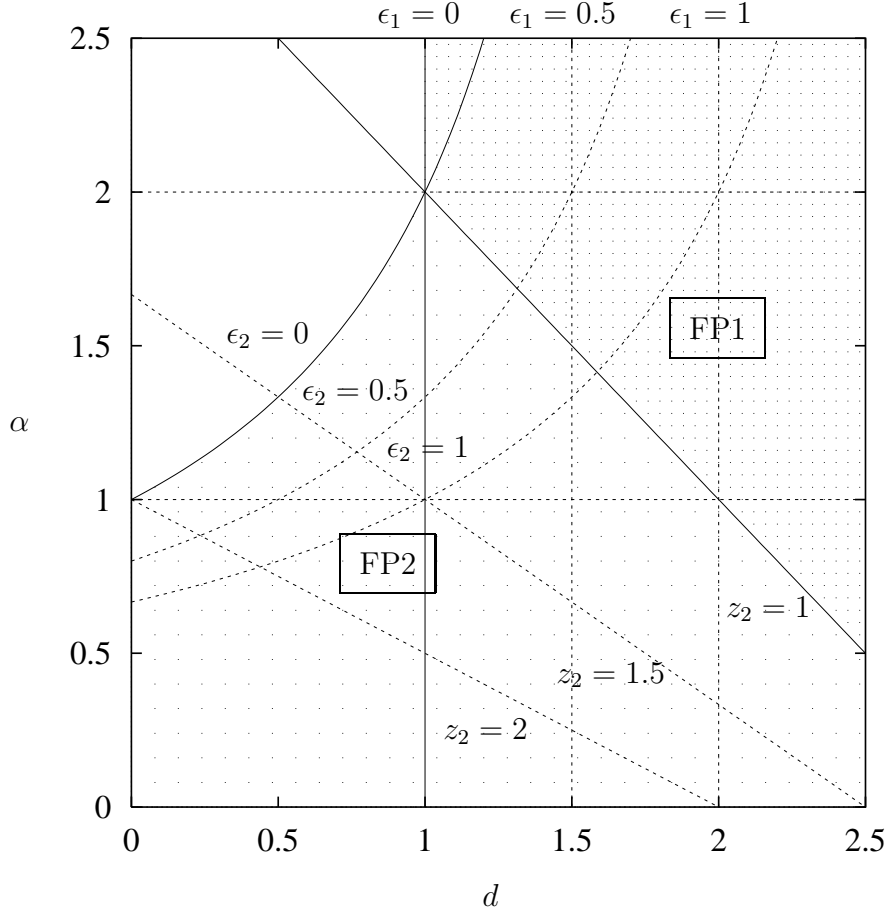,width=100\unitlength}\begin{picture}(0,0)  
\put(-26,55){\fbox{FP1}}\put(-61,30){\fbox{FP2}}  
\put(-57,88){\makebox(0,0)[bl]{$\epsilon_1=0$}}  
\put(-43,88){\makebox(0,0)[bl]{$\epsilon_1=0.5$}}  
\put(-26,88){\makebox(0,0)[bl]{$\epsilon_1=1$}}  
\put(-73,54){\makebox(0,0)[bl]{$\epsilon_2=0$}}  
\put(-65,48){\makebox(0,0)[bl]{$\epsilon_2=0.5$}}  
\put(-56,42){\makebox(0,0)[bl]{$\epsilon_2=1$}}  
\put(-20,25){\makebox(0,0)[bl]{$z_2=1$}}  
\put(-38,18){\makebox(0,0)[bl]{$z_2=1.5$}}  
\put(-50,12){\makebox(0,0)[bl]{$z_2=2$}}  
\put(-95,46){\makebox(0,0){$\alpha$}}  
\put(-45,-4){\makebox(0,0){$d$}}  
\end{picture}\par\bigskip\bigskip}  
  
\caption{The shaded areas FP1 and FP2 represent the regions of stability of  
the two fixed points in the $ (d,\,\alpha) $ plane
in the $N=3$ case.\label{stability}}  
\end{figure}  
  
We can now assess the stability of the two fixed points.  
In the \( (d,\, \alpha ) \) plane  
we distinguish the regions (see Fig. \ref{stability}
for the $N=3$ Heisenberg case):  
\begin{lyxlist}{00.00}  
  
\item [FP1:]\(\epsilon _{1}>0,\;   2N-3-d-(N-2)\alpha<0 \);  
  
\item [FP2:]\( \epsilon _{2}>0,\; 
  2N-3-d-(N-2)\alpha>0,\) i.e. \( z_{2}>1 \).  
  
\end{lyxlist}  
In region FP1 there exists only the corresponding fixed point, which  
 is stable. In region FP2 there appears a second, stable fixed point,  
while the point FP1 becomes unstable. The crossing between the  
regions FP1 and FP2 takes place for 
\( 2N-3-d-(N-2)\alpha=0 \).  
Notice that on the border line 
%in region FP2 
one gets  
\( \eta _{1}=\eta _{2}=2-\alpha  \) and  
\( z_{2}=z_{1}=1 \), so that when moving from the  
region FP2 to FP1 the critical exponents  
\( \eta  \) and \( z \) vary \em continuously\em .  
  
\begin{figure}  
{\setlength{\unitlength}{0.009\textwidth}\centering  
\epsfig{file=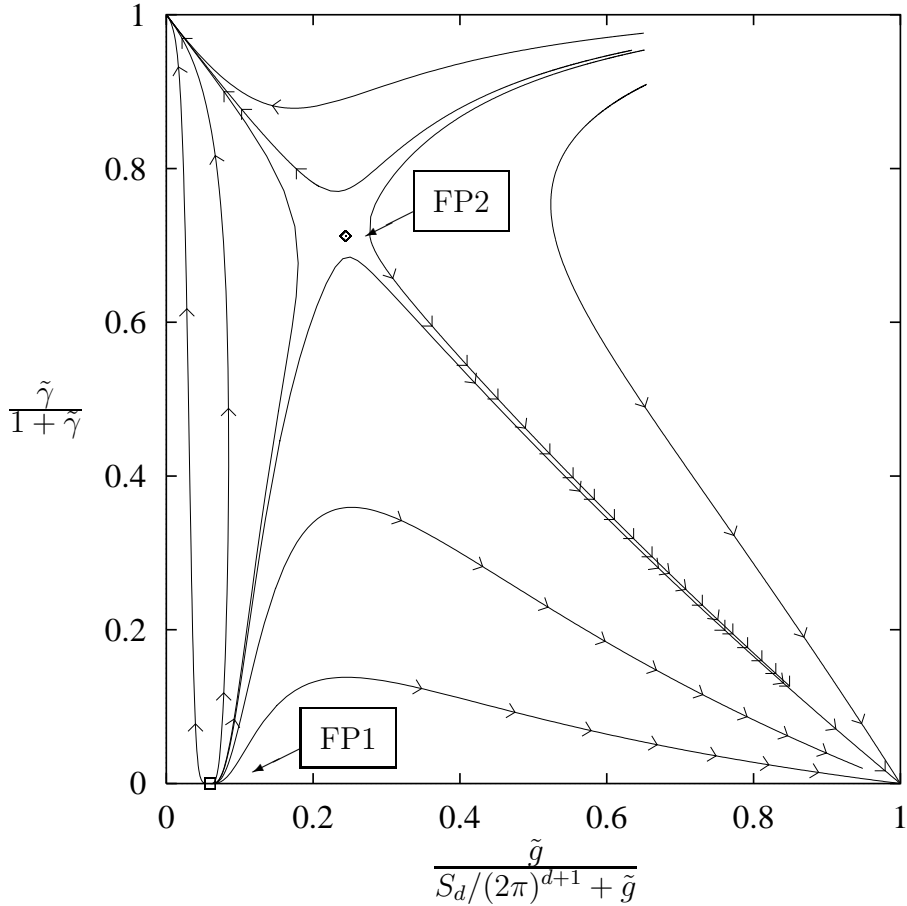,width=100\unitlength}  
\begin{picture}(0,0)\put(-97,45){\makebox(0,0){$  
\frac{\mbox{$\tilde\gamma$}}{\mbox{$1+  
\tilde\gamma$}}$}}\put(-45,-4){\makebox(0,0){$  
\frac{\mbox{$\tilde{g}$}}{\mbox{$S_d/(2\pi)^{d+1}+  
\tilde{g}$}}$}}\put(-70,9){\fbox{FP1}}  
\put(-70,9){\vector(-2,-1){5}}\put(-58,66){\fbox{FP2}}  
\put(-58,66){\vector(-2,-1){5}}\end{picture}\par\bigskip\bigskip\smallskip}

\caption{The RG flow projected on the compactified  
\protect\( (\tilde g ,\, \tilde\gamma )\protect \) plane  
(\protect\( d=1.2\protect \), 
\( \alpha =1\protect \),  
\( N=3\protect \)).  
\label{rgflow}}  
\end{figure}

Finally we report in Fig. \ref{rgflow} a numerical computation   
of the structure of the RG flow for typical values of the parameters.  
To this purpose it is convenient to introduce the rescaled variables  
\( \tilde\gamma =c^{\alpha }\gamma  \), \( \tilde g =cg \), so that  
(\ref{grt1_}-\ref{grt3_}) take the form  
\begin{eqnarray}  
\label{gra1} \frac{\d\tilde g }{\d l} & = & \{1-d+\tilde g  
[(N-2)\tilde I _{\alpha }(\tilde\gamma )-\tilde K _{\alpha }  
(\tilde\gamma )]\}\tilde g \\  
\label{gra2} \frac{\d\tilde\gamma }{\d l} & = & [2-\alpha  
-\tilde g (\tilde I _{\alpha }(\tilde\gamma )+\alpha  
\tilde K _{\alpha }(\tilde\gamma ))]\tilde\gamma \\  
\label{gra3} \frac{\d}{\d l}\frac{1}{c^{2}} & = &2 [1-z+  
\tilde g \tilde K _{\alpha }(\tilde\gamma )]\frac{1}{c^{2}}  
\end{eqnarray}  
where (\ref{gra1}) and (\ref{gra2}) are decoupled from (\ref{gra3})  
and do not contain the dynamic  
exponent \( z \). This means that in the new variables the RG trajectories  
project on the \( (\tilde g ,\, \tilde\gamma ) \) plane and the  
two lines of fixed points project on  
two corresponding fixed points. This fact was not \em a priori \em obvious  
and greatly facilitates the analytical and qualitative study of the  
RG trajectories.  
  
We notice the interesting fact that there exists  
a range of values \( \alpha_* \le \alpha <2 \) for which one obtains a  
\( z\equiv 1 \) dynamic scaling exponent  
in the proximity of the quantum critical point, even in the presence  
of a damping term. In other words, there can be ``soft'' damping terms  
which do not spoil the spin-wave behavior of the fixed point.  
Within the $\epsilon$-expansion for $N=3$
$\alpha_*$ is given by $\alpha_*=3-d$ (cf. Fig. \ref{stability}).
We shall argue about the extension of these results to the
\( d=2 \) case in the next subsection.
  
\subsection{Beyond the one-loop analysis}  
  
The above analysis of the stability of FP1 vs. FP2  
bears a strong similarity with the  
study of the \( \phi ^{4} \)  
theory with long range interactions \cite{fmn72,sak73}. For small  
\( \epsilon _{1} \) and \( \epsilon _{2} \) one has  
\( \alpha \simeq 2 \) and \(d \simeq 1 \)
and therefore \( \eta \simeq 2-\alpha  \),  
coherently with \cite{fmn72}. The  
two fixed points studied in \cite{sak73} correspond to our FP1 and  
FP2; in our approach they are selected by the value of  
\( z^{(0)} \)\em .  
\em The continuity of \( \eta  \) for \( \alpha \rightarrow 2 \)  
appears as a consequence of their relative  
stability, as in \cite{sak73}. The computations here are  
considerably simpler,  
since first-loop calculations are sufficient to establish the  
general picture.   
Another difference is in the presence of the dynamic exponent \( z \),  
which was absent in the  
classical works \cite{fmn72,sak73}.  
  
It is worth stressing the fact, noticed after Eq. (\ref{sviluppo}),  
that the non-analyticity  
of the term \( |\omega |^{\alpha } \) is not reproduced under  
renormalization, at least at  
first loop. As a consequence \( \gamma \) does not get singular  
corrections by itself, but only scales according to its bare  
dimension and to the anomalous field dimension\footnote{
More precisely, this discussion and Eq. (\ref{expgamma}) below
refer to the coupling \(\gamma/g\) of the dissipative term in 
Eq. (\ref{azione1}).
In the critical region \(g=g_*\) and the behavior
of \(\gamma\) coincides with the behavior of \(\gamma/g\).
}.
The indication coming from \cite{fmn72,sak73,aharony76}  
is that this  
will be true also at higher orders. This would imply that,  
as an exact result,   
 \( \gamma  \) scales in the critical region according to the exponent  
\begin{equation}  
\label{expgamma}  
x_{\gamma }=2-\alpha z-\eta   
\end{equation}  
irrespectively from the   
approximation. 
The value \( z_{2} \) can then be obtained directly from the  
condition \( x_{\gamma }=0 \), giving   
\begin{equation}  
\label{zeta}  
z_{2}=\frac{2-\eta }{\alpha }  
\end{equation}  
which is in agreement with our one-loop analysis and is  
expected to remain valid at higher  
loop orders as well.   
  
This exact relation can be used to settle the problem  
of the crossover from FP1 to FP2.  
This crossover takes place when  
\( z_{2}=z_{1}=1 \), that is, using  
(\ref{zeta}), when \( \alpha =\alpha_*\equiv 2-\eta(\alpha_*)\),
where \(\eta (\alpha)\) is the anomalous field dimension
for a given $\alpha$. Continuity of $\eta$ through 
the crossover implies \( \eta(\alpha_*)=\eta_1=\eta_2 \) with 
\( \eta_1 \) being  the critical exponent of the  
sigma model with zero damping in \( d+1 \).
For twodimensional systems, like the cuprates,  \( d+1=3 \).
Numerical estimates  for the critical indices of   
the \( \mathrm {O}(3) \) symmetric \(\phi^4\) field theory obtained from  
summed perturbation  
series at fixed dimension \( 3 \) (see \cite{lz77} and Table 25.4 of  
\cite{zinn89}) give \( \eta \simeq 0.033\pm 0.004 \)
and \( \alpha_* \simeq 1.966 \pm 0.004\).
In this respect the \( \epsilon\)-expansion greatly  
overestimates \( \eta \) and underestimates \( \alpha_*\)
in \( d=2 \) (see Fig. 1).  
This suggests that,
as a consequence of this small value of \( \eta \),  
since the relation (\ref{zeta}) is probably an exact relation, 
 for \( \alpha \) near one the physically  
relevant line of fixed point would be FP2.   
  
It appears therefore that the \( z=1 \) behavior observed in the cuprates  
can be accounted for in this framework only in the hypothesis that  
some mechanism like   
the reduction of the spectral weight of low-frequency quasi-particle  
is so strong to effectively produce a \( |\omega |^{\alpha } \)  
damping term with \( \alpha  \) very close  
to \( 2 \).

\section{The ordered phase: a scaling approach}  
  
\label{ordphase}

%\subsection{Form of the damping term }

In the ordered phase the effect of 
holes in the small-doping limit of the 
cuprates acts as a perturbation of the \(T=0\) 
antiferromagnetically ordered ground state.
The description of the interplay of spin and fermion
degrees of freedom 
when approaching
the QCP from the ordered side presents
major difficulties.
The simple introduction in the effective sigma model
of a \(\gamma|\omega|^\alpha\) term is not satisfactory, since
this term would be relevant also in the 
proximity of the ordered critical point \(g=0\), and
would completely
destroy 
the spin-wave
picture of
the ordered ground state.

In the ordered phase
the appearance of a non-zero
staggered magnetization \(N_0\) 
is accompanied by the presence of 
two quasi-particle bands,
separated by an energy \(\Delta\) 
(for a mean field analysis see \em e.g. \em Ref. \cite{scs95}).
As a consequence, one would expect 
the decay of spin-wave excitations 
in particle-hole pairs to be disfavored for
\(|\omega|\lesssim\Delta\), and an effective damping be felt only
for \(|\omega|\gtrsim\Delta\).
Moreover, 
in an AFI-PM transition
\(\Delta\) should vanish
at the QCP
as well as \(N_0\) does. 
Were this not the case, since \(\Delta=0\) in the disordered phase
one could hardly match the critical behavior on the two sides
of the transition.
We are not in a condition to directly link the scale \(\Delta\) to 
the staggered magnetization \(N_0\) since the precise relation, if any,
should emerge from the unknown microscopic dynamics of the 
quasiparticles\footnote{Notice for instance that in a large-U Hubbard model,
the size of the charge gap in the insulating phase is not only related
to magnetism, but is instead mainly induced by the strong
correlation}.
In this Section we 
try to get some insight 
in this difficult problem using simple
scaling arguments.
We will conjecture for \(\Delta\)
a scaling law with a critical exponent \(y\),
computable in principle 
from the full theory including also the fermionic 
degrees of freedom, 
and we shall try to assess the value of \(y\) 
in a self-consistent way,  once the value of $\alpha$ for the
damping in the disordered phase is given.
We emphasize that this procedure is not as reliable as the 
RG computation exposed in the preceding Sections
and has an heuristic character.

A model damping term which could take into account the appearance of the 
scale \(\Delta\) is
\begin{equation}
f(\omega)=\gamma\frac{|\omega|^b}{(\omega^2+\Delta^2)^{\frac{b-\alpha}{2}}}
\simeq\left\{
\begin{array}{ccc} 
r|\omega|^b & , & |\omega|<\Delta \\
\gamma|\omega|^\alpha & , & |\omega|>\Delta
\end{array}
\right.
\label{model1}
\end{equation}
with \(1\le\alpha<2\le b<3\), and 
\( r=\gamma\Delta^{\alpha-b} \).
We shall later assume that  
\(\Delta\) vanishes at the QCP, so that 
the crossover between the two regimes is continuously shifted 
while approaching 
the transition.
Here we consider \(\Delta \) as a free parameter at the same level 
as \(r\) and \(\gamma\).
Assuming that changing the scale, \(f(\omega)\) 
approximately maintains the form (\ref{model1}) with 
varying values of \(r\) and \(\gamma\)
one finds from (\ref{rhatf}) the characteristic
exponents
\begin{equation}
x_\gamma=2-\eta -\alpha z,\qquad x_r=2-\eta -bz,\qquad
x_\Delta=z.
\label{gammarcritexp}\end{equation}

Our phenomenological assumptions
suggest that damping be relevant only
at high frequencies.
In this regime 
the spin-wave velocity \(c_\mathrm{sw}\) goes as
\(c_\mathrm{sw}\sim s^{1-z_2}\)~\footnote{
Let us briefly comment about the relation 
of the physical spin-wave
velocity \(c_\mathrm{sw}\) with the running coupling constant
\(c\) and the dynamic exponent \(z\).
In the neighborhood of a critical point, \(z\) is determined
by the condition that \(c\) 
does not renormalize.
This corresponds to attributing all the renormalization of
the \(\omega^2/c^2\) term to the anomalous frequency scaling,
\(\omega\sim k^z\).
The running coupling constant \(c\) differs
from the physical spin-wave velocity \(c_\mathrm{sw}\), which is
measured at any scale \(s\) using a fixed system of physical units: one has
\(c_\mathrm{sw}\sim s^{1-z}c\).
So, although \(c\) does not renormalize by construction,
when \(z>1\) the physical velocity \(c_\mathrm{sw}\) 
vanishes at the transition.
In the case of the undamped sigma model the value
\(z=1\) is obtained as a consequence
of Lorentz invariance, so that there is no difference
between \(c\) and \(c_\mathrm{sw}\) and the latter remains finite
at the QCP.
However, the damping term breaks Lorentz invariance and can drive
\(c_\mathrm{sw}\) to zero at the QCP.
}. 
On the other hand, at any given scale the relevant \(\omega\)-modes
are characterized by the condition \(\omega^2=O(1)\).
So, damping is relevant up to the scale \(\bar s\) where \(\Delta\)
becomes of order \(1\).  Beyond this scale the gap for spin-wave decay
is visible and \(f(\omega)\) is irrelevant.
If \(\Delta\) at the ``microscopic" scale \(s=1\) 
had a finite value, eventually \(f(\omega)\)
would be always irrelevant. 
The simplest way to take into account the physical condition
that \(\Delta\) should vanish approaching the transition is to assume that
\begin{equation}
\left.\Delta\right|_{s=1}\sim (g_*-g_0)^{y}
\label{initcond}\end{equation}
where \(g_0= \left. g\right|_{s=1}\) and \(g_0\to g_*\).
This condition encodes in a simplified,
phenomenological way the complicated interaction
between damping and antiferromagnetic order, mediated in the full theory
by the fermionic degrees of freedom.

Taking into account (\ref{initcond}) and the scaling
exponents (\ref{gammarcritexp}) we get 
\begin{equation}
\bar s\sim(g_*-g_0)^{-y/z_2},\qquad g_0\to g_*
\label{barssimxi}\end{equation}
Approximately at this scale damping becomes irrelevant and the critical 
behavior starts being characterized by \(z=1\).
This gives \(c_\mathrm{sw}\) vanishing as
\begin{equation}
c_\mathrm{sw}\sim\bar s^{1-z_2}\sim(g_*-g_0)^{y(z_2-1)/z_2 },\qquad
g_0\rightarrow g_* 
\label{cswsimbars}\end{equation}

For \( \alpha \) not too close to two (specifically for \(\alpha <\alpha_*\))
by approaching the quantum transition from the disordered
phase one observes, as already discussed, a dissipative
behavior dominated by the fixed point FP2, with the scaling law
\(\omega\sim k^{z_2}\).
Deep into the broken symmetry phase one instead has a propagating
behavior, 
\(\omega\sim c_\mathrm{sw} k\). This should 
match the critical behavior at the scale
\(\xi\sim(g_0-g_*)^{-\nu}\),
with \(\nu\) the corresponding critical index.
Continuity of the critical behavior in
a neighborhood of the critical point imposes then
\(c_\mathrm{sw}\sim \xi^{1-z_2}\), {\it i.e.}
\[c_\mathrm{sw}\sim|g_*-g_0|^{\nu(z_2-1)},\qquad g_0\to g_*\]
This is compatible with (\ref{cswsimbars}) only for
\begin{equation}y=\nu z_2\label{yz2}\end{equation}
In terms of \( \xi \) this condition on the critical behavior of the
energy separation \(\Delta\) gives \(\delta \sim \xi^{-z_2} \).

\section{Conclusions}
\label{conclusions}
It is not unreasonable to expect that the damped sigma
model we considered contains some of the physics of
the quantum critical point of the underdoped layered
cuprates.
For this reason we performed a RG analysis directly on this
model, trying to maintain sufficient generality in the form
of the effective damping term, in particular introducing
an exponent \(\alpha \ge 1 \) that could account for the observed
pseudogap behavior. 
In the broken symmetry phase we tried to get some insight
on the interplay between damping and AF ordering using simple
scaling arguments.

In both the ordered and the disordered phases,
assuming continuity of the hydrodynamic behaviors at
the quantum transition, we obtained that for \(\alpha<\alpha_*\)
(where \(\alpha_*(d=2)\) is expected to be very close to 2)
it is not possible to account for
the observed \(z=1\) scaling law.
It was observed in Ref. \cite{scs95} that the \(z=1\)
scaling can be recovered at higher temperatures, when
typical frequencies would become larger than the \(T=0\)
value of the damping term.
However, experimentally it is the \(z=1\) behavior which
is observed at lower temperatures, while a relaxational
\(z=2\) behavior is observed at higher \(T\) \cite{cps96}.
It was argued \cite{mp94,cps96} that this could be due to
the neglected \(T\)-dependence of the damping term \(\gamma\).
We propose in this paper that the same effect can be accounted for 
by assuming that the decay of the quasi-particle spectral
weight for low frequencies produces a ``soft" damping
term \(\sim|\omega|^\alpha\) with \(\alpha\ge\alpha_*\).
The \(z=1\) scaling would then be a consequence
of an almost linear decay of the quasi-particle
spectral weight. 
We notice that this reduced weight is fully consistent
with the observed occurrence of a pseudogap below a crossover temperature
$T^*$ in the underdoped region of the phase diagram for the cuprates.
The origin of this pseudogap might arise from
extrinsic mechanisms like local Cooper-pair formation
and/or stripe fluctuations. Alternatively a pseudogap could be 
the very consequence of the 
intrinsecally non Fermi liquid nature of the 
system~\cite{andersonlutt} or of 
strong critical fluctuations occurring
near the QCP. 
In this last regard singular AF fluctuations have been shown 
within a spin-fermion model to strongly modify
the quasiparticles properties \cite{hlubina} in the proximity of the
``hot''points of the Fermi surface separated by
the AF wavevector. Since the particle-hole excitations
around these points are responsible for the low-energy spin-wave 
decay, this naturally affects the damping of spin excitations,
in a complicated interplay which deserves further investigation.
We notice that the fluctuation-induced 
spectral weight reduction for the
fermionic excitations is a definite signature of a
non-Fermi liquid metallic state. This is in agreement with
general arguments put forward by Anderson (see {\it e.g.},
Ref. \cite{ba97}) on the difficulty in continuously
connecting a charge-gapped insulating phase with
a normal Fermi-liquid phase. We remark, however, 
that Anderson's arguments extend the presence of a
non-Fermi-liquid phase all over the metallic state as a 
consequence of low dimensionality. On the other hand, 
within the QCP scenario, the violation of the Fermi-liquid 
properties is a consequence of critical fluctuations, which are
only present around the critical point. As soon as one moves away 
from this point, the fluctuations loose their critical character
thus providing simple perturbative corrections to the
Fermi liquid state.

It is worth noting that the analysis carried out in this paper
is not only related to the superconducting cuprates, but it has
connections with other physically interesting
problems, like the spin chains with long range
interaction and macroscopic quantum tunnelling.
By considering the model (\ref{azione1}) in \(d=0\),
it is seen that the frequency axis is left
as the only relevant ``direction'' (cf. the technical considerations
presented in Appendix \ref{compcritind}).
As a first consequence, all the critical behavior should be
reformulated in terms of frequency by suitably rescaling of the
critical indices by \(1/z\).
The frequency axis becomes then equivalent to a
single space direction, 
via the dynamic critical index \(z\),
thus establishing a connection with 
onedimensional classical models. 
Fourier transforming back to (imaginary) times 
the damping term in (\ref{azione1}) gives rise to a long range
interaction of the type $|t-t'|^{-1-\alpha}$. 
Our results can be directly extended to the case \(0<\alpha<1\).
This leads to the
comparison with the onedimensional (classical) 
$N$-component spin model  
with long-range interaction considered in Ref. \cite{kosterlitz}.
One can recognize that, once the spatial term is dropped in (\ref{azione1})
the connection is established through the coupling combinations
$\gamma/g$ and $gc^2$. This allows to rewrite the RG equation for
$g/\gamma$ in the same form
as Eq. (7) of Ref. \cite{kosterlitz}, with $g/\gamma$ (to be 
identified with $T$) having a fixed point (FP2 in our model)
and $(1/gc^2)_*=0$.

Whitin the $d=0$ case, $N=2$ deserves special attention because 
of the connection with  
macroscopic quantum tunnelling~\cite{caldeiralegget,variSC}.
This connection is realized by observing
that the $\mathrm{O}(2)$ non-linear $\sigma$ model can be rewritten in terms
of an angular degree of freedom and that the addition of the magnetic field 
produces a cos-like interaction (sine-Gordon model). In $d=0$
the model then describes a single degree of freedom with a kinetic
($\omega^2$) term in a multivalley potential. In the presence of
dissipation, a linear-in-frequency term can
appear, leading again to long-range interactions on the (imaginary) time axis.
Models of this kind were considered by Chakravarty \cite{chakravarty82} in the
case of a double well potential and successively by Schmid \cite{schmid}
to describe a dissipative quantum particle in a  onedimensional periodic
potential. In particular, the only difference between our Eq. 
(\ref{azione1}) at $N=2$ and this latter model 
is the non-periodicity of the dissipative term of
Ref. \cite{schmid}.

In the model of Schmid,
for vanishing $h$ ($g=0$ in his notation), a line of fixed points
for $\eta$ (corresponding to our $\gamma/g$) was found, 
with a critical point
$\eta_c$ separating a regime with relevant $h$ from one with irrelevant $h$. 
The long-range term introduced in \cite{schmid} is quadratic in the 
fields,
so that the model for $h=0$ is free and $\eta$ (i.e. $\gamma/g$)
does not renormalize\footnote{This is analogous to the situation found
in the classical $\mathrm{O}(2)$ non-linear $\sigma$ model in $d=2$ 
for $h=0$,
where a singular point appears at a special value of the 
coupling (see, for instance, \cite{zinn89})}.
A (dual) fixed point was also found for very large $h$, where the model of Ref.
\cite{schmid} was stated to be 
equivalent to the one of Ref. \cite{chakravarty82}.
We observe that, differently from the model of 
Ref. \cite{schmid} our long-range term
mantains the periodicity in the angular variable and 
provides an interaction. This renormalizes the coupling, thus 
allowing \(\gamma/g\) to flow under renormalization even at \(h=0\). 
One could conjecture that a new fixed point arises at finite \(h\).
An adequate analysis would require a RG study 
at finite \(h\) going beyond the \(\epsilon\)-expansion and
taking into account instantonic corrections.
This is an interesting open problem
which deserves further investigation.

{\bf Acknowledgments:} We gratefully acknowledge interesting
discussions with C. Di Castro, A. Chubukov and I. Kolokolov.

\appendix

\section{Computation of the critical indices}  
\label{compcritind}
  
For the computation of the eigenvalues \( \omega ^{\mathrm{R}},\,  
\omega ^{\mathrm{A}} \)  
of the linearization of the RG transformation (\ref{grt1_}-\ref{grt3_})  
in the point FP2 it is convenient  
to rewrite it in the form (\ref{gra1}-\ref{gra3}). The secular equation is  
found to   
depend on  
\begin{equation}  
\mathcal{I}_{\alpha }  =  \left. u\frac{d}{du}\log  
\tilde I _{\alpha }(u)\right| _{u=u_{*}},\quad  
\mathcal{K}_{\alpha }=\left. u\frac{d}{du}\log  
\tilde K _{\alpha }(u)\right| _{u=u_{*}}  
\label{iaka1}  
\end{equation}  
with $u=\tilde\gamma \equiv c^\alpha\gamma$ and 
$\tilde I _{\alpha }$ and $\tilde K _{\alpha }$ given after
Eq. (\ref{vincolo}).
For \( \epsilon =d-2+2/\alpha \rightarrow 0 \) one has  
\( u_{*}\rightarrow +\infty  \), and in this limit it  
is easy to derive from  
(\ref{loop},\ref{conv}), for \( \alpha >1 \):  
\begin{equation*}  
\tilde I _{\alpha }(u)  \simeq  
\frac{S_{d}}{(2\pi )^{d+1}}\frac{2\pi }{\alpha \sin  
\frac{\pi }{\alpha }}u^{-1/\alpha },~~~~  
\tilde K _{\alpha }(u)  \simeq  
\frac{S_{d}}{(2\pi )^{d+1}}\frac{\pi (\alpha -1)}{2\sin  
\frac{\pi }{\alpha }}u^{1/\alpha }  
\end{equation*}  
which substituted in (\ref{iaka1}) gives  
\begin{equation}  
\label{iaka}  
\mathcal{I}_{\alpha }\simeq -\frac{1}{\alpha },  
\quad \mathcal{K}_{\alpha }\simeq \frac{1}{\alpha }  
\end{equation}  
Solving (\ref{vincolo}) asymptotically  
for \( u_{*}\rightarrow +\infty  \) (\(\epsilon\rightarrow 0\))  
and fixed \( \gamma _{*} \) gives  
\begin{equation*}  
u_{*}^{2}  \simeq   
\left\{ \frac{4[(N-2)\alpha +1](2-\alpha )}  
{\alpha ^{3}(\alpha -1)}\right\} ^{\alpha }\frac{1}{  
\epsilon ^{\alpha }},~~~~  
c_{*}  \sim   \frac{1}{\sqrt{\epsilon }}  
\end{equation*}  
Using (\ref{iaka}) one easily finds   
\( \omega _{2}^{\mathrm{R}},\, \omega ^{\mathrm{A}}_{2} \) at   
first order in \(\epsilon\).  
  
For \( \alpha \rightarrow 1 \) the expressions (\ref{loop},\ref{conv})  
reduce to  
\begin{eqnarray}  
\label{loop2} I_{1}(c,\gamma ) & = &  
\frac{S_{d}}{(2\pi )^{d+1}}\left\{ \begin{array}{lll}  
\frac{4c}{\sqrt{4-(c\gamma )^{2}}}\arctan  
\frac{\sqrt{4-(c\gamma )^{2}}}{(c\gamma )}&, & 0<c\gamma <2\\  
\frac{2c}{\sqrt{(c\gamma )^{2}-4}}\log \frac{c\gamma +  
\sqrt{(c\gamma )^{2}-4}}{c\gamma -\sqrt{(c\gamma )^{2}-4}}&, & c\gamma >2  
\end{array}  
\right. \\  
 &  & \nonumber \\  
\label{conv2} K_{1}(c,\gamma ) & = & \frac{S_{d}}{(2\pi )^{d+1}}  
\frac{\gamma }{2}  
\end{eqnarray}  
The same expressions can  
be obtained by a direct evaluation of (\ref{loop0},  
\ref{kappaf}) for \( \alpha =1 \). In particular  
the original expression (\ref{loop0}) shows that  
\( I_{1} \) is in fact an analytic function  
of \( c \) and \( \gamma  \). However, some problems arise when   
using expressions (\ref{loop2}) and (\ref{conv2})
to compute  critical indices
directly for $\alpha=1$ instead of considering the $\alpha \to 1$
limit of the general expressions given in Section 3.1.
For \(\alpha=1\) in fact one expands around the   
critical dimension \( d_{\mathrm{c}}=0 \) and the  
factor \( S_{d} \) coming from angular  
integration vanishes. If  the same computation is performed using  
field-theoretical methods, instead of  
Wilson's integration on the momentum shell, no such problems arise:  
\em e.g. \em one has   
\begin{equation}  
\int \frac{\d \omega \d ^{d}k}{k^{2}+\omega ^{2}/c^{2}+  
\gamma \omega ^{\alpha }}  =  \frac{2}{2-\alpha }  
\frac{\pi /2}{\sin \frac{\pi d}{2}}\frac{2\pi ^{d/2}}  
{\Gamma (\frac{d}{2})}  
   \frac{\Gamma \left( \frac{1-d}{2-\alpha }\right)  
   \Gamma \left( \frac{\alpha \epsilon }{2(2-\alpha )}\right) }  
   {\Gamma \left( 1-\frac{d}{2}\right) }c^{  
   \frac{\alpha \epsilon }{2-\alpha }}\gamma ^{\frac{d-1}{2-\alpha }}  
\end{equation}  
which correctly behaves as \( \epsilon^{-1 } \) for  
\( \epsilon \rightarrow 0 \), independently  
of the value of \( \alpha  \). Notice that the $\Gamma^{-1}(d/2)$
factor coming from $S_d$ is cancelled by the factor
$\left( \sin(d\pi/2) \right)^{-1}$ and the $\epsilon^{-1}$ divergence
arises from $\Gamma(\alpha \epsilon /2(2-\alpha ))$.  
The problems with the  
Wilson method are resolved  
if one makes a more symmetric choice of the integration shell, \em e.g.  
\em integrating over all the \( \vec\pi _{\vec k ,\omega } \)  
with \( e^{-l}<\sqrt{k^{2}+\omega ^{2}}<1 \) (which implies introducing  
a frequency cutoff). With this choice \( I_{\alpha } \) is  
finite for \( d\rightarrow 0 \). We observed  
that \( I_{\alpha },\, K_{\alpha } \) computed with either  
choice of cutoffs have the same \( c,\, \gamma  \) dependence,  
the only effect of the choice of the symmetric shell being that the  
integrals do not vanish for \( d\rightarrow 0 \).   
In particular, the expressions for \(\omega_2^\mathrm{R}\),  
\(\omega_2^\mathrm{A}\) given in the text are valid for  
\(1\le\alpha<2\): as a matter of fact,  
they are computed from the values (\ref{iaka}), which  
do not contain the angular integration factor  \(S_d\).  

\end{document}